# Mapping the Human Body at Cellular Resolution -- The NIH Common Fund Human BioMolecular Atlas Program

Short Title: The Human BioMolecular Atlas Program (HuBMAP)

One Sentence Summary: HuBMAP supports technology development, data acquisition, and spatial analyses to generate comprehensive, molecular and cellular 3D tissue maps.

HuBMAP Consortium±

Author list can be found in the Appendix.

Contact Author: Michael Snyder, Alway M344, 300 Pasteur Drive, School of Medicine, Stanford University; Tel: (650) 736-8099; Email: mpsnyder@stanford.edu

Word count (2868), Abstract/Preface (100), Figures (3), References (51)

±Author list:
*Corresponding author

**Writing Group**
Michael P. Snyder[1*], Shin Lin[2*], Amanda Posgai[3*], Mark Atkinson[3*], Aviv Regev[4,5], Jennifer Rood[4], Orit Rozenblatt-Rosen[4], Leslie Gaffney[4], Anna Hupalowska[4], Rahul Satija[6,7], Nils Gehlenborg[8], Jay Shendure[9], Julia Laskin[10], Pehr Harbury[11], Nicholas A. Nystrom[12], Ziv Bar-Joseph[13], Kun Zhang[14], Katy Börner[15], Yiing Lin[16], Richard Conroy[17], Dena Procaccini[17], Ananda L. Roy[17], Ajay Pillai[18], Marishka Brown[19], Zorina S. Galis[19]

1. Department of Genetics, Stanford School of Medicine, Stanford, CA 94305
2. Department of Medicine, University of Washington, Seattle, WA 98195
3. Department of Pathology, University of Florida Diabetes Institute, Gainesville, FL 32610
4. Klarman Cell Observatory Broad Institute of MIT and Harvard, Cambridge MA 02142
5. Howard Hughes Medical Institute, Koch Institute of Integrative Cancer Research, Department of Biology, Massachusetts Institute of Technology, Cambridge MA 02140
6. New York Genome Center, New York City, NY 10013
7. New York University, New York NY 10012
8. Department of Biomedical Informatics, Harvard Medical School, Boston, MA 02115
9. Department of Genome Sciences, University of Washington, Seattle, WA 98105
10. Department of Chemistry, Purdue University, West Lafayette, IN 47907
11. Department of Biochemistry, Stanford University School of Medicine, Stanford, CA 94305
12. Pittsburgh Supercomputing Center, Carnegie Mellon University, Pittsburgh, PA 15213
13. Department of Computational Biology, School of Computer Science, Carnegie Mellon University, Pittsburgh, PA 15213
14. Department of Bioengineering, University of California San Diego, La Jolla, CA 92093
15. Department of Intelligent Systems Engineering, School of Informatics, Computing, and Engineering, Indiana University, Bloomington, IN 47408
16. Department of Surgery, Washington University School of Medicine, St Louis, MO 63110
17. Division of Program Coordination, Planning, and Strategic Initiatives, National Institutes of Health, Bethesda, MD 20892

18. National Human Genome Research Institute, National Institutes of Health, Bethesda, MD 20892
19. National Heart, Lung, and Blood Institute, National Institutes of Health, Bethesda, MD 20892

**HuBMAP Tissue Mapping Centers (TMCs)**
**Caltech-UW TMC**
Long Cai[20]*, Jay Shendure[9], Cole Trapnell[9], Shin Lin[2], Dana Jackson[9]

20. Department of Biology and Biological Engineering, California Institute of Technology, Pasadena CA 91125

**Stanford-WashU TMC**
Michael P. Snyder[1]*, Garry Nolan[20], William James Greenleaf[1], Yiing Lin[16], Sylvia Plevritis[22], Sara Ahadi[1,], Hayan Lee[1], Stephanie Nevins[1], Ed Esplin[1], Aaron Horning[1], Amir Bahmani[1]

21. Department of Microbiology, Stanford School of Medicine, Stanford, CA 94305
22. Department of Radiology, Stanford School of Medicine, Stanford, CA 94305

**UCSD TMC**
Kun Zhang[14]*, Xin Sun[14], Sanjay Jain[23], James Hagood[24], Peter Kharchenko[8]

23. Department of Medicine, Washington University in St. Louis, St. Louis, MO 63110
24. Department of Pediatrics, University of North Carolina School of Medicine, Chapel Hill, NC 27599

**University of Florida TMC**
Mark Atkinson[3]*, Bernd Bodenmiller[25], Todd Brusko[3], Michael Clare-Salzler[3], Harry Nick[26], Kevin Otto[27], Amanda Posgai[3], Clive Wasserfall[3], Sergio Maffioletti[25]

25. Institute of Molecular Life Sciences, University of Zurich, CH-8057 Zurich, Switzerland
26. Department of Neuroscience, University of Florida, Gainesville, FL 32611
27. Department of Biomedical Engineering, University of Florida, Gainesville, FL 32610

**Vanderbilt TMC**
Richard M. Caprioli[28]*, Jeffrey M. Spraggins[28]*, Danielle Gutierrez[28], Nathan Heath Patterson[28], Elizabeth K. Neumann[28], Raymond Harris[29], Mark deCaestecker[29], Raf Van de Plas[30], Ken Lau[31]

28. Mass Spectrometry Research Center, Department of Biochemistry, Vanderbilt University, Nashville, TN 37235
29. Department of Medicine, Vanderbilt University School of Medicine, Nashville, TN 37232
30. Delft Center for Systems and Control, Delft University of Technology, 2628 CD Delft, The Netherlands.
31. Department of Cell and Developmental Biology, Vanderbilt University, Nashville, TN 37232

**Transformative Technology Development Groups (TTDs)**
**California Institute of Technology TTD**
Long Cai[20]*, Guo-Cheng Yuan[32], Qian Zhu[32], Ruben Dries[32]

32. Department of Biostatistics and Computational Biology, Dana-Farber Cancer Institute, Boston. MA 02215

**Harvard TTD**
Peng Yin[33, 34]*, Sinem K. Saka[33, 34], Jocelyn Y. Kishi[33, 34], Yu Wang[33, 34], Isabel Goldaracena[33, 34]

33. Wyss Institute for Biologically Inspired Engineering, Harvard University, Boston, MA 02115
34. Department of Systems Biology, Harvard Medical School, Boston, MA 02115

**Purdue TTD**
Julia Laskin[10]*, DongHye Ye[10, 35], Kristin E. Burnum-Johnson[36], Paul D. Piehowski[36], Charles Ansong[36], Ying Zhu[36]

35. Department of Electrical and Computer Engineering, Opus College of Engineering, Marquette University, Milwaukee, WI 53233
36. Biological Sciences Division, Pacific Northwest National Laboratory, Richland, WA 99352

**Stanford TTD**
Pehr Harbury[11]*, Tushar Desai[37], Jay Mulye[11], Peter Chou[11], Monica Nagendran[37]

37. Department of Internal Medicine, Division of Pulmonary & Critical Care, Stanford University School of Medicine, Stanford, CA 94305

**HuBMAP Integration, Visualization, and Engagement (HIVE) Collaboratory**

**Carnegie-Mellon - Tools Component**
Ziv Bar-Joseph[13]*, Sarah A. Teichmann[38]*, Benedict Paten[39]*, Robert F. Murphy[13], Jian Ma[13], Vladimir Yu. Kiselev[38], Carl Kingsford[13], Allyson Ricarte[13], Matthew Ruffalo[13]

38. Cellular Genetics Programme, Wellcome Sanger Institute, Hinxton, Cambridgeshire, CB10 1SA, UK
39. Department of Biomolecular Engineering, Jack Baskin School of Engineering, University of California Santa Cruz, Santa Cruz, CA 95064

**Harvard Medical School - Tools Component**
Nils Gehlenborg[8]*, Peter Kharchenko[8], Margaret Vella[8], Chuck McCallum[8]

**Indiana University Bloomington - Mapping Component**
Katy Börner[15]*, Leonard E. Cross[15], Samuel H. Friedman[40], Randy Heiland[15], Bruce Herr II[15], Paul Macklin[15], Ellen M. Quardokus[15], Lisel Record[15], James P. Sluka[15], Griffin M. Weber[8]

40. Opto-Knowledge Systems, Inc., Torrance, CA 90502

**University of Pittsburgh - Infrastructure Component**
Nicholas A. Nystrom[12]*, Jonathan C. Silverstein[41]*, Philip D. Blood[12], Alexander J. Ropelewski[12], William E. Shirey[41]

41. Department of Biomedical Informatics, University of Pittsburgh, Pittsburgh, PA 15206

**University of South Dakota - Collaboration Core**
Paula Mabee[42]*, W. Christopher Lenhardt[43], Kimberly Robasky[43, 44, 45], Stavros Michailidis[46]

42. Department of Biology, University of South Dakota, Vermillion, SD 57069
43. Renaissance Computing Institute, University of North Carolina, Chapel Hill, NC 27517
44. Department of Genetics, University of North Carolina, Chapel Hill, NC 27517
45. School of Information and Library Science, University of North Carolina, Chapel Hill, NC 27517
46. Knowinnovation Inc., Buffalo, NY, 14209

**New York Genome Center - Mapping Component**

Rahul Satija*[6, 7], John Marioni[38, 47, 48], Aviv Regev[4, 5], Andrew Butler[6, 7], Tim Stuart[6, 7], Eyal Fisher[48], Shila Ghazanfar[48], Jennifer Rood[48], Leslie Gaffney[48], Gokcen Ersalan[48], Tommaso Biancalani[4]

47. European Molecular Biology Laboratory, European Bioinformatics Institute (EMBL-EBI), Wellcome Genome Campus, Hinxton, CB10 1SD, United Kingdom
48. Cancer Research UK Cambridge Institute, University of Cambridge, Li Ka Shing Centre, Robinson Way, Cambridge, CB2 0RE, United Kingdom

**NIH HuBMAP Working Group**

Richard Conroy[17], Dena Procaccini[17], Ananda Roy[17], Ajay Pillai[18], Marishka Brown[19], Zorina Galis[19], Pothur Srinivas[19], Aaron Pawlyk[49], Salvatore Sechi[49], Elizabeth Wilder[17], James Anderson[17]

49. National Institute of National Institute of Diabetes and Digestive and Kidney Diseases, National Institutes of Health, Bethesda, MD 20892

Contact PIs for respective TMCs, TTDs, or HIVE are listed first.

**Transformative technologies are enabling the construction of three dimensional (3D) maps of tissues with unprecedented spatial and molecular resolution. Over the next seven years, the NIH Common Fund Human Biomolecular Atlas Program (HuBMAP) intends to develop a widely accessible framework for comprehensively mapping the human body at single-cell resolution by supporting technology development, data acquisition, and detailed spatial mapping. HuBMAP will integrate its efforts with other funding agencies, programs, consortia, and the biomedical research community at large towards the shared vision of a comprehensive, accessible 3D molecular and cellular atlas of the human body, in health and various disease settings.**

**Introduction**

The human body is an incredible machine. Trillions of cells, organized across an array of spatial scales and a multitude of functional states, contribute to a symphony of physiology. While we broadly know how cells are organized in most tissues, a comprehensive understanding of the cellular and molecular states and interactive networks resident in the tissues and organs, from organizational and functional perspectives, is lacking. The specific 3D organization of different cell types, together with the effect of cell-cell and cell-matrix interactions in their natural milieu, have a profound impact on the normal function, natural aging, tissue remodeling, and disease progression in different tissues and organs. Recently, new technologies have enabled the molecular characterization of a multitude of cell types[1–4] and mapping of their spatial relationships in complex tissues at unprecedented scale and single cell resolution. These advancements create an opportunity to build a high-resolution atlas of 3D maps of human tissues and organs.

HuBMAP (see https://commonfund.nih.gov/hubmap) is a NIH-sponsored program with the goals of developing an open framework and technologies for mapping the human body at cellular resolution as well as generating foundational maps for several tissues obtained from normal individuals across a wide range of ages. Whereas GTEx[5] is a previous NIH sponsored project that examined DNA variants and bulk tissue expression patterns across approximately a thousand individuals, HuBMAP is a distinct project focused on generating molecular maps spatially resolved at the single cell level but on a more limited number of subjects. To achieve these goals, HuBMAP has been designed as a cohesive and collaborative organization, having a culture of openness and sharing using team science-based approaches[6]. The HuBMAP Consortium (https://hubmapconsortium.org/) will actively work with other ongoing initiatives including the Human Cell Atlas[7], Human Protein Atlas[8], LIfeTime[9], and related NIH-funded consortia mapping specific organs (including brain[10], lungs[11], kidney[12], and genitourinary[13] regions) and tissues (especially pre-cancer and tumors[14]), as well as other emerging programs.

**HuBMAP organization and approaches**

The HuBMAP consortium ([https://hubmapconsortium.org/](https://hubmapconsortium.org/)) is comprised of members with a broad diversity of expertise (e.g., molecular, cellular, developmental, and computational biologists, measurement experts, clinicians, pathologists, anatomists, biomedical and software engineers, and computer and data information scientists) and is organized into three components: 1) Tissue Mapping Centers (TMCs), 2) HuBMAP Integration, Visualization & Engagement (HIVE) collaborative components, and 3) Innovative Technologies Groups (TTDs and RTIs) (**Fig. 1**). Throughout the program, HuBMAP will grow the range of tissues and technologies studied through a series of funding opportunities that have been designed to be synergistic with other NIH and international efforts. In the later stages of HuBMAP, demonstration projects will be added to show the utility of the generated resources and importantly, engage the wider research community to analyze HuBMAP data alongside data from other programs or from their own labs.

**Tissue and data generation**

The HuBMAP TMCs will collect and analyze a broad range of largely normal tissues, representing both sexes, different ethnicities and a variety of ages across the adult lifespan. These tissues (**Fig. 2**) include: 1) discrete, complex organs (kidney, ureter, bladder, lung, breast, colon); 2) distributed organ systems (vasculature); and 3) systems comprised of dynamic or motile cell types with distinct microenvironments (lymphatic organs: spleen, thymus, and lymph nodes). Tissue collection will occur at precisely defined anatomical locations (when possible, photographically recorded) according to established protocols that preserve tissue quality and minimize degradation. Beyond meeting standard regulatory requirements, to the greatest extent possible, samples will be consented for open access data sharing (i.e. public access without approval by data committees) to maximize their usage by the biomedical community.

To achieve spatially-resolved, single-cell maps, the TMCs will employ a complementary, iterative, two-step approach (**Fig. 3**). First, 'omic assays, which are extremely efficient in data acquisition, will be used to generate global genome sequence and gene expression profiles of dissociated single cells/nuclei in a massively parallel manner. The molecular state of each cell will be revealed by single cell transcriptomic[15] and in many cases chromatin accessibility[16,17] assays; imputation of transcription factor binding regions from the open chromatin data combined with the gene expression data will be used to explain the regulation of gene expression across the distinct cell types[18]. Second, spatial information (abundance, identities, and localization) will be acquired from various biomolecules (RNA[19], protein[20], metabolites, and lipids) in tissue sections or blocks, using imaging methodologies such as fluorescent microscopy (confocal, multiphoton, lightsheet, and expansion), sequential Fluorescence In Situ Hybridization [seqFISH[21,22]]), imaging mass spectrometry[23,24], and imaging mass cytometry (IMC[25–28]). The extensive single cell/nucleus profiles obtained will inform *in situ* modalities (e.g., single cell/nucleus RNA-seq will be used to choose probes to RNA or proteins), which provide spatial information for up to hundreds of molecular targets of interest. These data will allow for computational registration of cell-specific epigenomic/transcriptomic profiles to cells on a histologic slide to reveal various microenvironmental states. They will potentially include information about protein localization to cytoplasm, nucleus, or cell surface; phosphorylation; complex assembly; extracellular environment; and cellular phenotype determined by protein marker coexpression. Registration and computational integration of complex imaging data will provide biological insight beyond any single imaging mode[23,29]. The powerful combination of single cell profiling and multiplexed *in situ* imaging will provide a pipeline for constructing multi-omics spatial maps for the various human organs and their cellular interactions at a molecular level.

The TMCs will apply complementary methods for data collection with an emphasis on processes to ensure the generation of high quality data and standardized metadata annotations. Benchmarking, quality assurance and control (QA/QC) standards, and standard operating procedures (SOPs), where appropriate, will be developed for each stage of the methodological process and be made available to promote rigor, reproducibility and transparency. It is expected that QA/QC standards for both biospecimens and data will evolve as tissue collection, processing techniques, storage/shipping conditions, assays, and data processing tools change, and as HuBMAP interacts and collaborates with other related efforts, as they have for other Consortium projects[30–35]. Where possible, metadata related to preanalytical variables (e.g. annotations and nomenclature) and technologies will be harmonized and protocols and standards will be shared with the wider research community.

**Computational approaches for building an integrated tissue map across scales**

The diversity of data generated by HuBMAP, ranging across macro- and microscopic scales (e.g. anatomic, histologic, cellular, molecular and genomic), and multiple individuals is essential to its core mission. Exploring each of these valuable datasets collectively will yield an integrated view of the human body. Hence, HuBMAP will develop analytical and visualization tools bridging spatial and molecular relationships in order to help generate a high-resolution 3D molecular atlas of the human body.

The volume of data generated and collected by HuBMAP will require the utilization, extension and development of tools and pipelines for data processing. While we expect that initial data processing tools will be based on methods developed by consortium members, HuBMAP will also work with and incorporate algorithms developed by other programs and the wider research community to supplement, enhance or update its pipelines. To this end, HuBMAP will develop one or more portals tailored to emerging use cases identified through a series of user needs. These open source portals will use recognized standards and be interoperable with other platforms, such as the HCA Data Coordination Platform (DCP), allowing to readily add, update, and use new software modules (e.g. as with Dockstore[36] and Toil[37]). The portion of HuBMAP data that will be open source can live on or be accessed from multiple platforms, enhancing its utility. This infrastructure will enable external developers to apply their codes, applications, open application programming interfaces (APIs), and data schema to facilitate customized processing and analysis of HuBMAP data in concert with other data sources. Furthermore, by actively working with other global and NIH initiatives, the Consortium will seek to reduce the barriers to browsing, searching, aggregating, and analyzing data across portals and platforms.

To fully integrate spatial and molecular data across individuals, HuBMAP will create a common coordinate framework (CCF) that defines a 3D spatial representation, leveraging both an early Consortium-wide effort to standardize technologies and assays using a single common tissue and the broader range of tissues of the human body analyzed across multiple scales (whole body to single cells). This spatial representation will serve as an addressable scaffold for all HuBMAP data, enabling unified interactive exploration and visualization (search, filter, details on demand) and facilitating comparative analysis across individuals, technologies, and labs[38,39]. To achieve these objectives, HuBMAP envisions a strategy inspired by other tissue atlas efforts[40–42] leveraging the identification of "landmark" features including key anatomical structures and canonical components of tissue organization (e.g., epidermal boundaries and normally spatially invariant vasculature) that can be identified in all individuals. These landmarks will enable a "semi-supervised" strategy for aligning and assembling an integrated reference, upon which HuBMAP investigators can impose diverse coordinate systems, including relative representations and zone-based projections. As one example, an open-source, computational histology topography cytometry analysis toolbox (histoCAT[43]) currently facilitates 2D and soon, 3D reconstruction. Ontology-based frameworks will be explored in parallel to effectively categorize, navigate, and name multi-scale data; synergies are expected between these two approaches. Whenever available, medical imaging, such as CT and MRI information, will help serve as a basis for landmarking and constructing the CCF.

**Technology development and implementation**

Quantitative imaging of different classes of biomolecules in the same tissue sample with high spatial resolution, sensitivity, specificity, and throughput is central to the development of detailed tissue maps. Although no single technique can fully address this challenge at present, the development and subsequent multiplexing of complementary capabilities provides a promising approach for accelerating tissue mapping efforts. The HuBMAP Innovation Technologies groups aim to develop several innovative approaches to address limitations of existing state-of-the-art techniques. For example, transformative technologies such as Signal Amplification by Exchange Reaction (SABER)[44,45], SeqFISH[22,46,47], and Lumiphore probes[48] will be refined to improve multiplexing, sensitivity, and throughput for imaging RNA and proteins across multiple tissues. Furthermore, new mass spectrometry imaging techniques will enable quantitative mapping of hundreds of lipids, metabolites, and proteins from the same tissue section with high spatial resolution and sensitivity[49,50]. There is also scope within the program to develop and test new technologies. These efforts will benefit from development of novel computational tools and machine learning algorithms, optimized first from data generated from a common tissue during the pilot phase, for data integration across modalities.

**Challenges**

Optimizing collection, preservation, and processing of a wide diversity of tissue types from multiple donors has been approached by previous programs such as GTEx[5]. However, one of the goals of HuBMAP, to generate comprehensive, interactive high resolution maps using a wide variety of assays, introduces an added level of complexity. Mapping functionally important biomolecules, including some of which we may not even be aware and for which sensitive, specific, and high-throughput assays are still lacking, will require devoted attention. Moreover, the volume and diversity of datasets are heretofore unprecedented for comprehensive data capture, management, mining, modelling, and visual exploration and communication. Integration of data from different modalities is necessary to generate robust maps; it will be necessary to develop the corresponding analysis and interactive visualization tools necessary to ensure that the data and atlas are widely accessible to the entire life-sciences community. Finally, given the enormity of a human atlas, HuBMAP faces the challenges of prioritization of tissues and technologies, sampling across tissues and donors, and optimally synergizing its efforts with international efforts. Determining the number of cells, fields of view, and samples needed to capture rare cell types, states or tissue structures, is an important challenge, but can be tackled with adaptive power analyses, leveraging the growing amount of data available both within HuBMAP and from other consortia as well as individual groups.

**Resources and Community Engagement**

HuBMAP is an important part of the international mission to build a high resolution cellular and spatial map of the human body, and we are firmly committed to close collaboration and synergy with the aforementioned initiatives to build an easy-to-use platform and interoperable datasets that will accelerate realization of a high-resolution human atlas. Shared guiding principles around open data, tools, and access will enable collaborative and integrated analyses of data produced across diverse consortia. To achieve this synergy, HuBMAP and other consortia will work together to tackle common computational challenges, such as cellular annotation, through formal and informal gatherings focused on addressing these problems, planned joint benchmarking and hands-on jamborees and workshops. Another example of the potential for close collaboration is in the study of the colon; multiple projects funded by HuBMAP, the Human Tumor Atlas Network, and the Wellcome Trust will be complemented by projects funded by the Leona M. and Harry B. Helmsley Charitable Trust. With projects focusing on partly distinct regions and diseases (e.g., normal tissue, colon cancer, and Crohn's disease) it will be important for all the programs to ensure data is collected and made available in a consistent manner, and HuBMAP will play an active role in such efforts. As a concrete next step, HuBMAP, in collaboration with other NIH programs, plans to hold a joint meeting with the Human Cell Atlas initiative to identify and work on areas of harmonization and collaboration during the spring of 2020. In parallel, HubMAP participants engage in meetings and activities of other consortia, such as the Human Cell Atlas or the Human Tumor Atlas Network, thus forming tight connections. We have started a series of open meetings to develop the CCF, with the first of these recently held in collaboration with the Kidney Precision Medicine Program and focused on the kidney.

HuBMAP will provide capabilities for data submission, access, and analysis following FAIR (Findable, Accessible, Interoperable, and Reusable) data principles[51]. We will develop policies for prompt and regular data releases in commonly-used formats, consistent with similar initiatives. We anticipate a first round of data release in the summer of 2020 with subsequent releases at timely intervals thereafter. Robust metadata will be comprised of all aspects of labeling and provenance including de-identified donor information (both demographic and clinical), details of tissue processing and protocols, data levels, and processing pipelines.

Indeed, engagement and outreach to the broader scientific community and other mapping centers is central to ensure that resources generated by HuBMAP will be leveraged broadly for sustained impact. To ensure that browsers and visualization tools from HuBMAP are valuable, the Consortium will work closely with anatomists, pathologists, as well as visualization and user experience experts; such as those having virtual or augmented reality expertise. As described earlier, it is expected that the diversity of normal samples included in this project will facilitate valuable comparative analyses, pinpointing how cells and tissue structures vary across individuals, throughout the lifespan, and in the emergence of dysfunction and disease. The program will build its resources with these use cases in mind and provide future opportunities, such as the demonstration projects, for close collaboration with domain experts. We also anticipate these data will be highly useful for new biomedical hypothesis generation, tissue engineering, developing robust simulations of spatiotemporal interactions and machine learning of tissue features, as well as for educational purposes.

**Conclusions**

Analogous to the release of the first human genome build, we anticipate the first reference 3D tissue maps will represent the "tip of the iceberg" in terms of their ultimate scope and eventual impact. HuBMAP, working closely with other initiatives, aspires to help build a foundation by generating a high resolution atlas of key organs in the normal human body and capturing inter-individual differences as well as acting as a key resource for new contributions in the growing fields of tissue biology and cellular ecosystems. Given the focus of HuBMAP on spatial molecular mapping, the Consortium will contribute to the community of efforts seeking similar goals, with a special emphasis on providing leadership to the development of analytical methods for its data types and for developing a common coordinate framework to integrate data. Ultimately, we hope to catalyze novel views on the organization of tissues, not only regarding which types of cells are neighboring one another, but the gene and protein expression patterns that define these cells, their phenotypes, as well as functional interactions. In addition to encouraging the establishment of intra- and extra-Consortium collaborations that align with HuBMAP's overall mission, the Consortium envisions an easily accessible, publicly available user-interface where data can be used to visualize molecular landscapes at the single cell level, pathways and networks for molecules of interest, and spatial and temporal changes across a given cell type of interest. Researchers will also be able to browse, search, download, and analyze the data in standard formats with rich metadata which, over time, will enable users to query and analyze datasets across similar programs.

Importantly, we envision the project's compilation of different types of multiomic information at the single cell level in a spatially-resolved manner to represent an important step in the advancement of our understanding of human biology and precision medicine. These data have the potential to re-define cell types/subtypes and their relationships within and between tissues beyond traditional understanding made available from standard methods (e.g., microscopy, flow cytometry). We hope this work will be part of a foundation that enables diagnostic interrogation, modeling, navigation, and targeted therapeutic interventions at such an unprecedented resolution so as to be transformative for the biomedical field.

**Acknowledgements**

This research is supported by the NIH Common Fund, through the Office of Strategic Coordination/Office of the NIH Director under awards OT2OD026663, OT2OD026671, OT2OD026673, OT2OD026675, OT2OD026677, OT2OD026682, U54AI142766, U54DK120058, U54HG010426, U54HL145608, U54HL145611, UG3HL145593, UG3HL145600, UG3HL145609, and UG3HL145623.

subpopulations by combining scRNAseq and sequential fluorescence in situ hybridization data. *Nat. Biotechnol.* (2018). doi:10.1038/nbt.4260

48. Cho, U. *et al.* Ultrasensitive optical imaging with lanthanide lumiphores. *Nat. Chem. Biol.* **14**, 15–21 (2018).

49. Yin, R. *et al.* High Spatial Resolution Imaging of Mouse Pancreatic Islets Using Nanospray Desorption Electrospray Ionization Mass Spectrometry. *Anal. Chem.* **90**, 6548–6555 (2018).

50. Zhu, Y. *et al.* Nanodroplet processing platform for deep and quantitative proteome profiling of 10-100 mammalian cells. *Nat. Commun.* **9**, 882 (2018).

51. Wilkinson, M. D. *et al.* The FAIR Guiding Principles for scientific data management and stewardship. *Sci Data* **3**, 160018 (2016).

**Figure Legends**

**Figure 1. The HubMAP consortium.**

The Tissue Mapping Centers (TMC) will collect tissues and generate spatially resolved, single cell data. Groups involved in Transformative Technology Development (TTD) and Rapid Technology Implementation (RTI) initiatives will develop emerging and more developed technologies, respectively, which, in later years, will be implemented at scale. Data from all groups will be rendered useable for the biomedical community by the HIVE. The groups will closely collaborate to iteratively refine the atlas as it is gradually realized.

**Figure 2. Key tissues and organs initially analyzed by the consortium.**

Using innovative, production-grade (i.e. "shovel ready") technologies, HuBMAP Tissue Mapping Centers (TMC) will generate data for single cell, 3D maps of various human tissues. In parallel, Transformative Technology Development (TTD) projects, and later Rapid Technology Implementation projects will refine assays and analysis tools on a largely distinct set of human tissues. Samples from individuals of both sexes and across different ages will be studied. The range of tissues will be expanded throughout the program.

**Figure 3. Map generation and assembly across cellular and spatial scales.**

HuBMAP aims to produce an atlas in which users can refer to a histologic slide from a specific part of an organ and in any given cell understand its contents on multiple 'omic levels--genomic, epigenomic, transcriptomic, proteomic, and/or metabolomic. To achieve these ends, centers will apply a combination of imaging, 'omics and mass spectrometry techniques to specimens collected in a reproducible manner from specific sites in the body. These data will be then be integrated to arrive at at high-resolution, high-content 3D map for any given tissue. To ensure inter-individual differences will not be confounded with collection heterogeneity, a robust common coordinate framework will be developed.

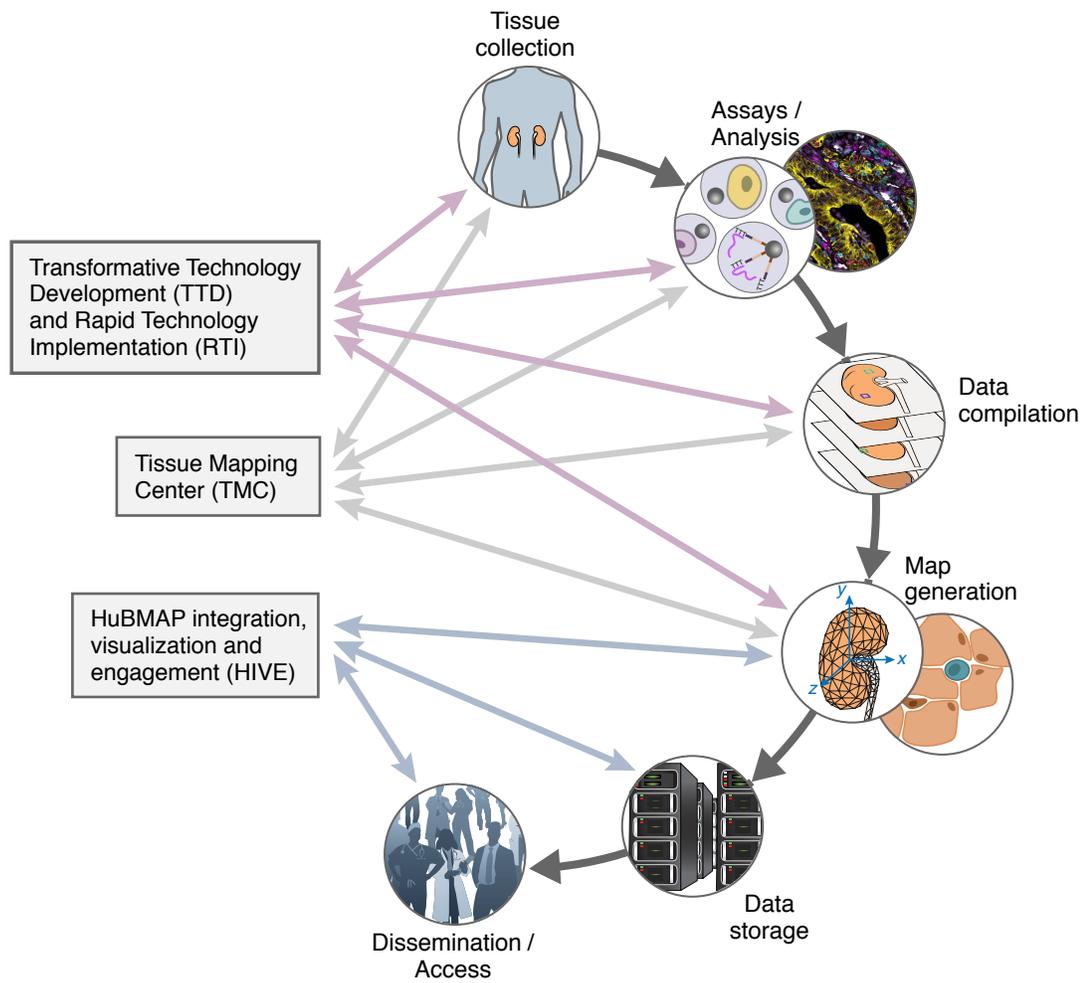

Figure 2

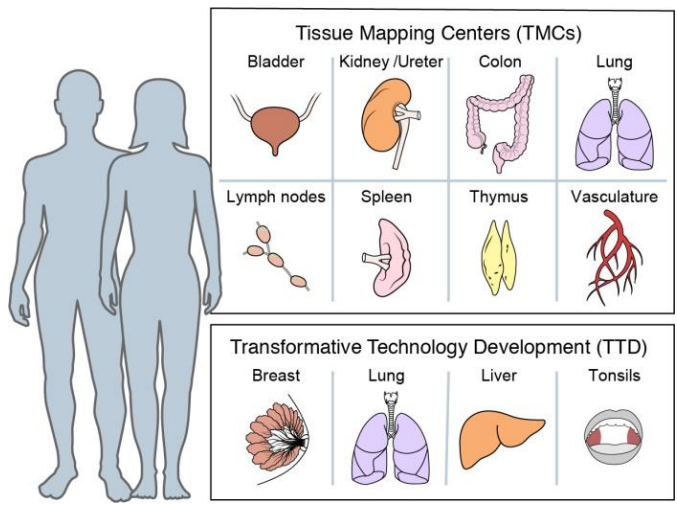

Figure 3

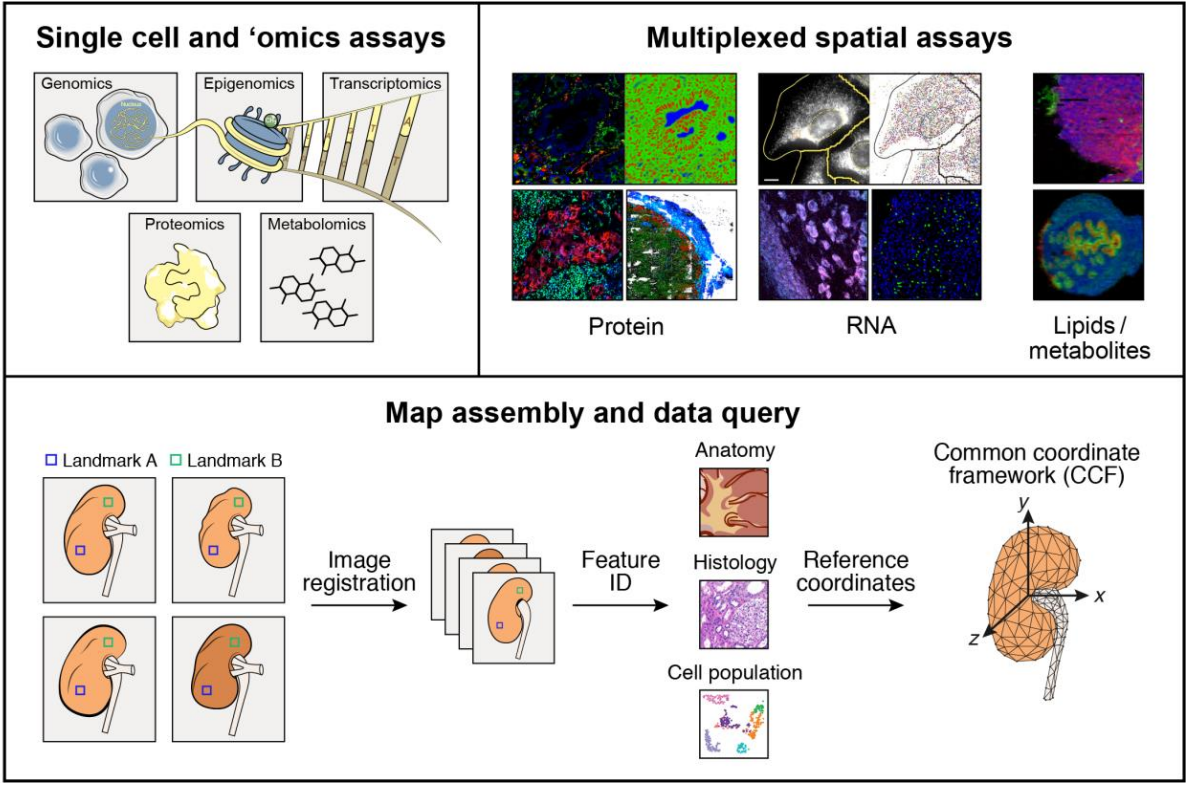